\documentclass[twocolumn]{aastex63}
\usepackage{amsmath}	% Advanced maths commands

\received{June 1, 2019}
\revised{January 10, 2019}
\accepted{\today}

\submitjournal{AJ}

\shorttitle{Neutron Star Merger Host Galaxy Mass Function Prediction}
\shortauthors{Rose et al.}

\graphicspath{{./}{figures/}}

\begin{document}

\title{Where binary neutron stars merge: predictions from IllustrisTNG}

\correspondingauthor{Jonah Rose}
\email{j.rose@ufl.edu}

\author[0000-0002-2628-0237]{Jonah C. Rose}
\affiliation{Department of Astronomy, University of Florida, Gainesville, FL 32611, USA}

\author[0000-0002-5653-0786]{Paul Torrey}
\affiliation{Department of Astronomy, University of Florida, Gainesville, FL 32611, USA}

\author[0000-0002-4832-0420]{K.H. Lee}
\affiliation{Department of Physics, University of Florida, Gainesville, FL 32611, USA}

\author[0000-0001-5607-3637]{I. Bartos}
\affiliation{Department of Physics, University of Florida, Gainesville, FL 32611, USA}

\begin{abstract}

The rate and location of Binary Neutron Star (BNS) mergers are determined by a combination of the star formation history and the Delay Time Distribution (DTD) function.
In this paper, we couple the star formation rate histories (SFRHs) from the IllustrisTNG model to a series of varied assumptions for the BNS DTD to make predictions for the BNS merger host galaxy mass function.
These predictions offer two outcomes:  
(i) in the near term: influence BNS merger event follow-up strategy by scrutinizing where most BNS merger events are expected to occur 
and 
(ii) in the long term: constrain the DTD for BNS merger events once the host galaxy mass function is observationally well determined.
From our fiducial model analysis, we predict that 50\% of BNS mergers will occur in host galaxies with stellar mass between $10^{10} - 10^{11}$ $M_\odot$, 68\% between $4 \times 10^{9} - 3\times 10^{11}$ $M_\odot$, and 95\% between $4 \times 10^8 - 2 \times 10^{12}$ $M_\odot$. 
We find that the details of the DTD employed does not have a strong effect on the peak of the host mass function. 
However, varying the DTD provides enough spread that the true DTD can be determined from enough electromagnetic observations of BNS mergers.
Knowing the true DTD can help us determine the prevalence of BNS systems formed through highly eccentric and short separation fast-merging channels and can constrain the dominant source of r-process material.

\end{abstract}


\section{Introduction}

Since the first discovery of gravitational waves by LIGO \citep{2015CQGra..32g4001L}, a growing number of compact object mergers have been detected. To date, two detections have been confirmed as binary neutron star (BNS) mergers \citep{Abbott2017,2020ApJ...892L...3A}. 
Of these two, the BNS event GW170817 was detected across the electromagnetic spectrum \citep{Abbott2017a}, beginning the age of multi-messenger astronomy. 

Within the next few years we expect tens of new BNS mergers events to be observed by LIGO, Virgo \citep{Acernese_2014} and KAGRA \citep{kagra2019}, broadening our understanding of BNS systems and their host galaxies \cite{2018LRR....21....3A}.
Developing a clearer understanding of the link between BNS merger events and their host galaxies is useful for multiple reasons in both the short and long term.

In the short term, having clear predictions for the BNS host galaxy mass function could inform follow-up strategies for future BNS merger events detected by LIGO/Virgo/KAGRA.
Locating the electromagnetic counterpart of GW events is difficult owing to narrowly peaked observability windows and large localization areas \citep{Smart2017, 2012Metzger}. 
GW localizations can extend tens-to-hundreds of square degrees, making them impractical to completely cover in a reasonable time after the initial event with most telescopes  \citep{2016Gehrels,Bartos_2013,2015ApJ...801L...1B,2018MNRAS.477..639B,2019MNRAS.490.3476B}. 
Long-term radio emission could allow sufficient time for follow-up observations, but this will only be possible for nearby events within dense circum-merger media \citep{2019MNRAS.485.4150B,2020arXiv200700563L,2020arXiv200805330G}.

GW follow-up strategies have taken two approaches to search the localization area more efficiently: covering the entire area or targeting galaxies \citep[e.g.][]{2014Bartos, 2017Arcavi, Chan2018, 2020Antier}.
Covering the entire localization area increases the likelihood of imaging the correct host galaxy, but risks missing the transient owing to the limited exposure times. 
\cite{2017Soares} used this method to successfully locate the kilonova after GW170817 by covering 80.7\% of the probability weighted localization area. 
In contrast, targeted follow-ups use galaxy catalogs to preferentially search galaxies based on select criteria (e.g. galaxy blue luminosity), potentially reducing the required number of pointings by a factor of 10 to 100 
and increasing exposure times \citep{2016Gehrels, 2020Ducoin}. 
This strategy was used in the first successful detection of the optical counterpart of GW170817 \citep{2017Coulter}.
However, it is more likely that targeted strategies will miss the event if the BNS merger takes place in a less-massive galaxy.
Efficient follow-up strategies are important for maximising the chance of identifying the electromagnetic counterpart with limited observations. 
One way to achieve this is to build a clearer understanding of the expected host galaxy mass function for BNS mergers.

Longer term, the link between BNS merger events and their host galaxies can help determine the dominant formation channel of $r$-process material by constraining the true BNS Delay Time Distribution (DTD) 
\citep{2016Marchant, 2018Barrett, 2019Mapelli, 2020Santoliquido, 2020McCarthy}.
Specifically, while core-collapse supernovae (SNe) and BNS mergers have been proposed as $r$-process formation channels,
the dominant $r$-process production channel must be able to recreate the observed decreasing trend in Eu/Fe vs Fe/H \citep{2015Matteucci}.
BNS mergers must produce $r$-process elements in less than 100\,Myr to dominate $r$-process element production \citep{2018Hotokezaka, 2017Cote}, and in less than 1\,Myr -- with a steep cutoff slope --  to be the source of all $r$-process material \citep{2015Matteucci}.
While it is possible for BNS systems to merge in this time \citep{2019SafarzadehDwarf}, 
they require highly eccentric orbits from high-velocity kicks or low initial separation from case BB mass transfer, both of which may not occur in BNS formation \citep{2017Tauris}. 
These models also predict a shallower DTD based on the current understanding of BNS formation channels \citep{2019Giacobbo, 2019Simonetti, Safarzadeh2019}.

Current stellar populations synthesis models suggest BNS DTDs are best modelled by power law distributions with an exponent between -1 and -1.5 ~\citep{2019Simonetti} and the minimum time from creation of the binary system to merger ($t_{\mathrm{min}}$) between 1\,Myr to 1\,Gyr \citep[e.g.][]{2019Simonetti, 2019SafarzadehDwarf}.
These models assert that the main formation channel that forms BNS systems begin with two OB stars that are close enough to undergo mass transfer  \citep{2017Tauris, 2018Giacobbo, 2019SafarzadehDwarf}. 
The rest of the systems are born through so-called fast-merging channels where a binary system forms 
with either a high eccentricity through large natal kicks, or with a small initial separation through unstable case BB mass transfer \citep{2017Tauris}.
If the BNS DTD could be observationally constrained, it would not only shed light on the physical formation channels (fast-merging vs. OB star mass transfer), but could also help constrain 
progenitor metallicity, 
common-envelope efficiency, 
natal kicks, 
mass ratio, 
and initial binary separation through comparisons with population synthesis codes \citep{2018Giacobbo, 2018Belczynski, 2012Dominik}.
Constraining the BNS DTD may be one of the best and most practical ways to constrain the physical origin and implications of BNS merger events.

Taken over a whole galaxy, the rate of BNS merger events can be determined by convolving the DTD with the star formation rate history (SFRH).
Metallicity is also accounted for in some models, but has been found to play a minor role in influencing the DTD \citep{2019Mapelli, 2019Giacobbo, 2017Cote, 2018Giacobbo}.
Previous studies have measured the BNS merger rate using SFRHs derived from EAGLE and Illustris cosmological simulations \citep{2019ArtaleMass, Mapelli2018, 2019Mapelli}, the FIRE zoom-in simulation \citep{2018Lamberts}, dark matter only simulations \citep{2019Marassi, 2018Cao}, or from semi-analytical models \citep{2020Adhikari, 2019Toffano} with population synthesis codes or an assumed DTD. 
Each of these models provide a different SFRH for the galaxies in that simulation which provide different distributions for BNS mergers given the same DTD. 

In this paper, we use the IllustrisTNG simulations to make predictions for the BNS merger host galaxy mass function. 
This extends the work that which has been presented in~\citep{Mapelli2018} and ~\citep{2019ArtaleMass} by focusing on the uncertainty/variation introduced by variations in the assumed DTD.
Moreover, we consider here how in the future an observed BNS host galaxy mass function could be used to constrain the real/underlying DTD.
To do this, we take as input the IllustrisTNG galactic SFRHs -- which are known to match a wide range of observed galaxy properties and galaxy scaling relations -- and employ varied assumptions about the BNS DTD.  
Our chosen SFRHs and DTDs allow us to demonstrate the galactic masses at which we expect most BNS mergers to occur, as well as to identify the level of variation that would be induced based on changes to the BNS DTD. 

The structure of this paper is as follows.  
In Section~\S\ref{sec:Methods} we outline our methods including a brief description of the IllustrisTNG simulations, our adopted DTDs, and our methods for calculating the galaxy-by-galaxy BNS merger rate.  
In \S\ref{sec:Results} we present our main results including predictions for the BNS merger host galaxy mass function and the sensitivity of this prediction to the assumed DTD.
In \S\ref{sec:Discussion} we discuss the implications of our results.
Finally, in \S\ref{sec:Conclusions} we discuss our results and conclude.

\section{Methods}
\label{sec:Methods}

In this paper, we make predictions for the BNS merger host galaxy mass function by adopting SFRHs from cosmological simulations and DTDs from basic stellar population synthesis models.

\subsection{Delay Time Distributions}
Generally, the current BNS merger rate for any galaxy is given by convolving its SFRH with the appropriate DTD.
The BNS merger rate for any collection of material (e.g. galaxy, volume, etc.) is given by
\begin{equation}
    r(t) = \int_0 ^t \psi(\tau) \Gamma(t - \tau, \, Z) d\tau
\end{equation}
where $\psi$ is the star formation rate, $\Gamma(t - \tau, \, Z)$ is the DTD, and the integration is performed from the Big Bang ($t=0$) to the time of observation, $t$~\citep[e.g.][]{Maoz2012}. The metallicity dependence, $Z$, in $\Gamma$ is only present in some DTDs, otherwise the DTD takes the form $\Gamma(t - \tau)$.
In the case of cosmological galaxy formation simulations, this can be reduced to a sum over contributions from all relevant stellar populations
\begin{equation}
    \label{eq:dtd}
    r(t) = \sum_j M_j  \Gamma (t - t_j, \,Z_j)
\end{equation}
where the sum is performed over all stars (or stellar populations) in the region of interest (e.g. within a specific galaxy), 
$M_j$ is the mass of each stellar particle or stellar population, 
$t_j$ is birth time of that stellar population such that $t-t_j$ is the age of the stellar population, 
and $Z_j$ is the metallicity of the stellar population.
For any cosmological galaxy formation simulation, the BNS merger rate is easily evaluated once a BNS DTD is specified.

\begin{figure*}
	\includegraphics[width=\columnwidth]{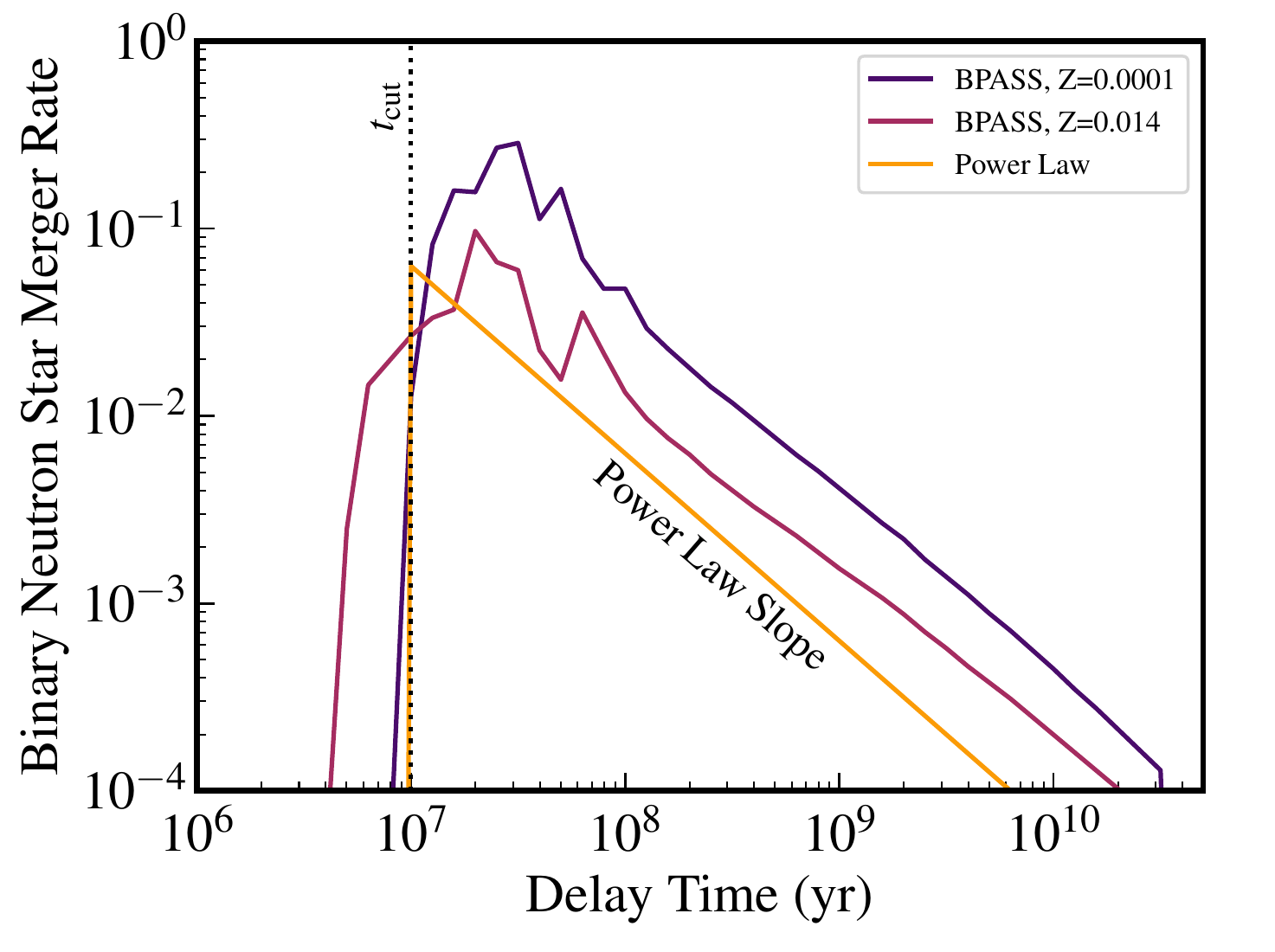}
	\includegraphics[width=\columnwidth]{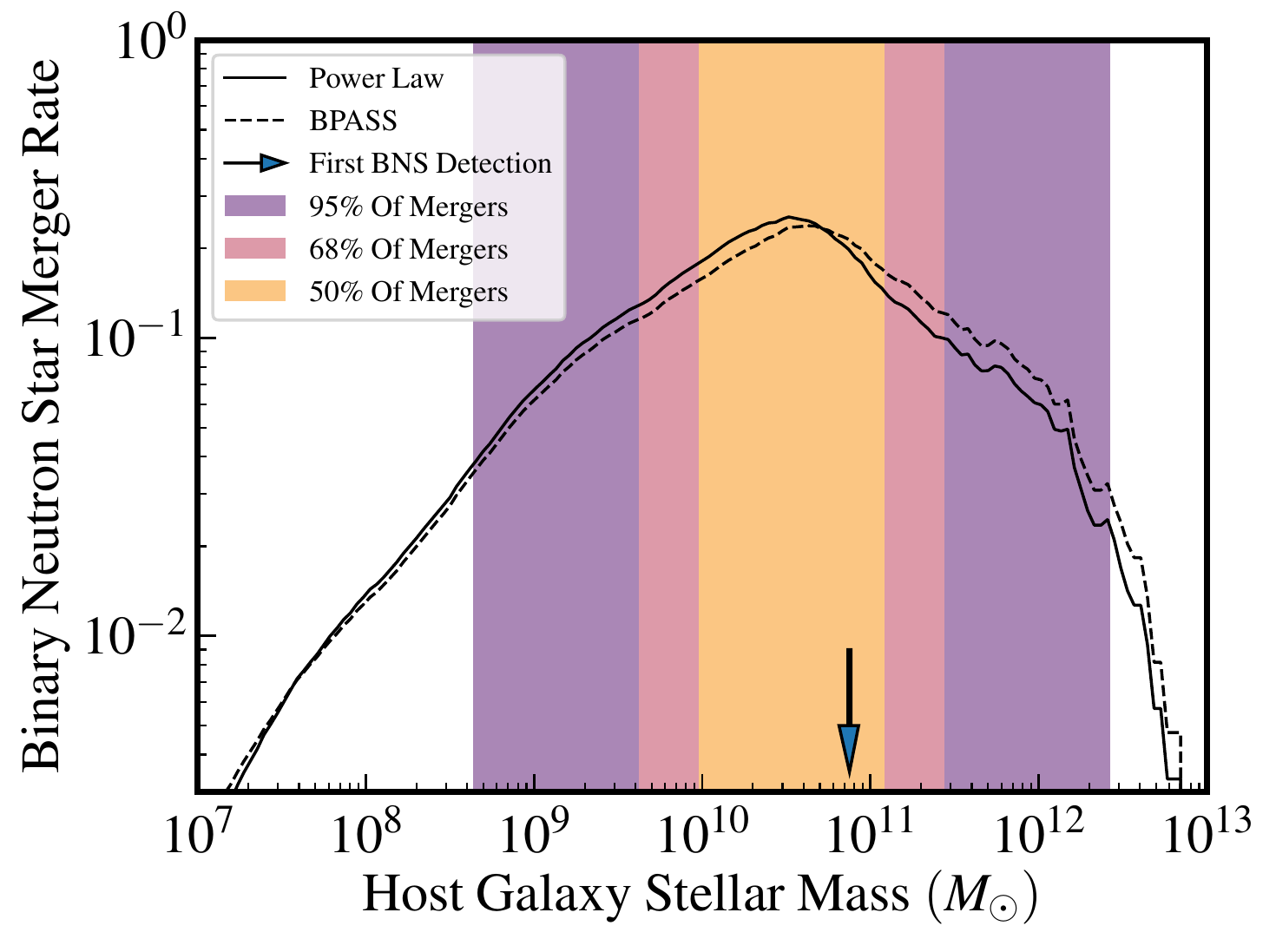}
    \caption{\textit{(left)} The two fiducial DTDs shown as the BNS merger rate vs time. The BPASS DTD is split into two lines to show the range covered by the DTD as the metallicity of the host star changes. \textit{(right)} The normalized BNS merger rate for the two fiducial DTDs as a function of the host galaxy stellar mass. The solid line shows the merger rate given the power law DTD with an exponent of s=-1 and $t_{\rm cut}$=0.01Gyr. The dashed line shows the merger rate given the BPASS DTD. Both merger rates have been normalized individually such that the total merger rate across the simulation for the given DTD is unity. The shaded bands show the mass range which contain 50, 68, and 95 percent of the mergers around the peak merger rate for the fiducial power law DTD. The arrow points to the host galaxy mass of the only BNS merger with a detected electromagnetic counterpart so far.}
    \label{fig:fiducial}
\end{figure*}

\subsection{Power Law DTD}
In this paper, we adopt two fiducial DTDs: 
(i) a simple paramaterized power law and 
(ii) the metalicity dependent DTD from BPASS~\citep{Eldridge2016}. 
The power law DTD is given by
\begin{equation}
    \Gamma (t-t_j) =
    \begin{cases} 
      0 & t-t_j \leq t_{\mathrm{cut}} \\
      \Gamma_0 (t-t_j)^s & t-t_j > t_{\mathrm{cut}} 
   \end{cases}
\end{equation}
where $\Gamma_0$ is a normalization coefficient, $s$ is the power law index, and $t_{\mathrm{cut}}$ is the minimum time/age before the first BNS merger event occurs.
Figure~\ref{fig:fiducial} shows our fiducial power law DTD ($s=-1$; $t_{\mathrm{cut}} =  10^7 \, \mathrm{yrs}$).
In addition to our fiducial power law DTD, we also consider DTDs that have varied power law exponents ranging from $s=-2$ to $s=2$ and cutoff times ranging from $t_{\mathrm{cut}}=0.001 \, \mathrm{Gyr}$ to $t_{\mathrm{cut}}=10 \, \mathrm{Gyr}$.

\subsection{BPASS DTD}

In addition to using a simple power law DTD, we adopt our second fiducial DTD from BPASS ~\citep{Eldridge2016} as shown in Figure~\ref{fig:fiducial}.
BPASS is a stellar population synthesis code which follows the evolution of a large suite of varying binary stars \citep{Eldridge2008, Eldridge2016}. 
The critical feature of BPASS that is important for this paper is that the simulated stellar population matches the observed population of binaries in abundance along with supernovae progenitors and rates \citep{Eldridge2008, Eldridge2013, Eldridge2015, Eldridge2016}. The remnants of supernova can have a mass in the full range between .1 and 300 $M_\odot$, allowing for more realistic evolution of these systems. 
The simulation also encompass a wide range of stellar metalicities which has been shown to potentially affect (albeit weakly) the DTD of BNS mergers \citep[e.g.][]{Mapelli2018}. 
The final time for the BNSs to merge in BPASS is then the sum of the progenitor stars' evolution and the in-spiral time once the BNS system has formed \citep{Eldridge2016}.
The tabulated BPASS BNS merger rates are a function of stellar population age and metallicity, $\Gamma(t-t_j, Z_j)$, which can be employed in conjunction with Equation~\ref{eq:dtd} to determine the total BNS merger rate.

%These paragraphs are not new, but were just moved around a bit.
We note that the overall normalization for all of our DTDs ($\Gamma_0$, in the case of the powerlaw DTD) can be specified to match the expected global rate of BNS mergers.
However, its exact value is not important to the present work as we are only interested in the relative/normalized distribution of BNS merger events as a function of galaxy mass.
We therefore normalize the total BNS merger rate across the entire simulation box to unity for each DTD individually. 
To achieve this, we divide the rate for an individual galaxy, $j$, by the rate of the entire box for the given DTD, $\Gamma$. The normalized form of Equation \ref{eq:dtd} becomes

\begin{equation}
    \label{eq:dtd_norm}
    R_i(t) = \frac{r_i(t)}{\sum_k M_k  \Gamma (t - t_k, \, Z_k) }
\end{equation}
where $r_i$ represents the BNS merger rate for an individual galaxy given by equation \ref{eq:dtd}, and the denominator sums over the rates of all galaxies ($k$) in the simulation box.

We linearly interpolate across the ages and metallicities presented in \cite{Eldridge2016}. 
We do not extrapolate outside of the provided metallicity values; all stars with $Z \leq 0.0001$ follow the DTD for $Z=0.0001$ and all stars with $Z \geq 0.014$ follow the DTD for $Z=0.014$. 
The BNS merger rate is then calculated using Equation \ref{eq:dtd_norm}.
The BPASS DTD includes information on natal kicks from the initial supernovae. For each supernova, the kick velocity and direction are determined from \cite{Hobbs2005}. For more information, see \cite{Eldridge2011}. There are no natal kicks included in the calculation of the power law delay time distributions, including the fiducial power law model.  
Thus, the power law DTDs are fully specified with two parameters controlling 
(i) the time of BNS mergers onset and 
(ii) the subsequent BNS merger rate evolution.

\subsection{The IllustrisTNG Simulation Suite}

In order to calculate the BNS merger rates, we adopt SFRHs from IllustrisTNG simulation \citep{Pillepich2018b, Nelson2018, Marinacci2018, Springel2018, Naiman2018}. 
IllustrisTNG is a suite of cosmological hydrodnamical simulations which includes a comprehensive galaxy formation model~\citep{Pillepich2018a, Weinberger2017} and builds upon the original Illustris model~\citep{Vogelsberger2013, Torrey2014}.
The critical feature of the IllustrisTNG simulations for this paper is that the simulations have been shown to broadly reproduce the cosmic star formation rate density and the redshift-dependent galaxy stellar mass function \citep{Pillepich2018b, Springel2018, 2018Nelson}. 
These tests give confidence that both the global and galaxy-by-galaxy star formation histories produced by IllustrisTNG reasonably match those of the Universe.

While it is possible that the constraints for any given galaxy differ from a ``real'' galaxy, when averaged over a large enough sample the trends are well matched. 
Specifically, we  employ the TNG-100-1 simulation which includes a $100 \; \rm{Mpc}$ cubed volume with hundreds-of-thousands of galaxies with varied SFRHs and self-consistently evolved metallicity distributions.
All stellar particles are used in equation \ref{eq:dtd_norm} to calculate the normalization, but results are only presented for galaxies down to a stellar mass of $10^7 M_\odot$.
We use the full information from the simulated galaxy populations including the age and metallicity distribution to evaluate the BNS merger rate.

\section{Results}
\label{sec:Results}

Figure~\ref{fig:fiducial} shows the DTDs (left) and host galaxy mass function (right) for our two fiducial setups.
The DTDs follow the same general form with mergers being most likely shortly after $t_{\mathrm{cut}}$ then dropping with increasing time. The main difference comes from the metallicity dependence in the BPASS DTD. 
Owing to the similarities in the DTDs, the resulting BNS merger host galaxy mass functions are remarkably similar.
In particular, we find that the peak of BNS mergers occurs in host galaxies with stellar masses around $M_* = 5 \times 10^{10}$ $M_\odot$ and that half of the mergers occur in galaxies with masses between $10^{10} M_\odot < M_* < 10^{11}$ $M_\odot$. 
Despite the significant added complexity of the BPASS models, the predicted BNS host galaxy mass function is not significantly different from the power law DTD.  Additionally, the mass of the host galaxy from GW170817~\citep{Blanchard2017} is indicated with a downward facing arrow in the right panel of Figure~\ref{fig:fiducial}.
While it is a sample size of one and should not be over-interpreted, we note that GW170817's host galaxy mass is in the peak region of expected BNS host galaxy masses for both of our fiducial DTD models.

\begin{figure*}
	\includegraphics[width=\columnwidth]{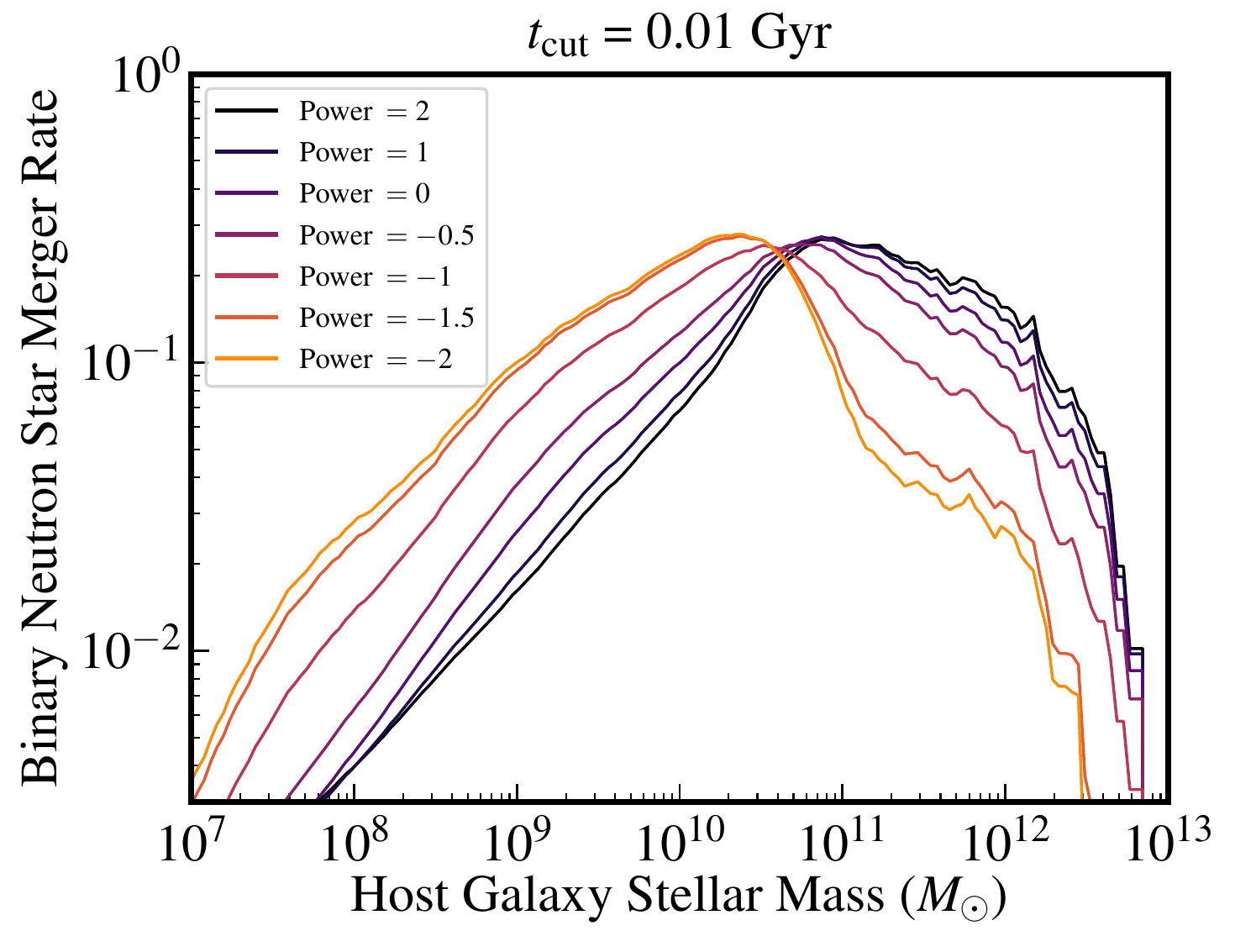}
	\includegraphics[width=\columnwidth]{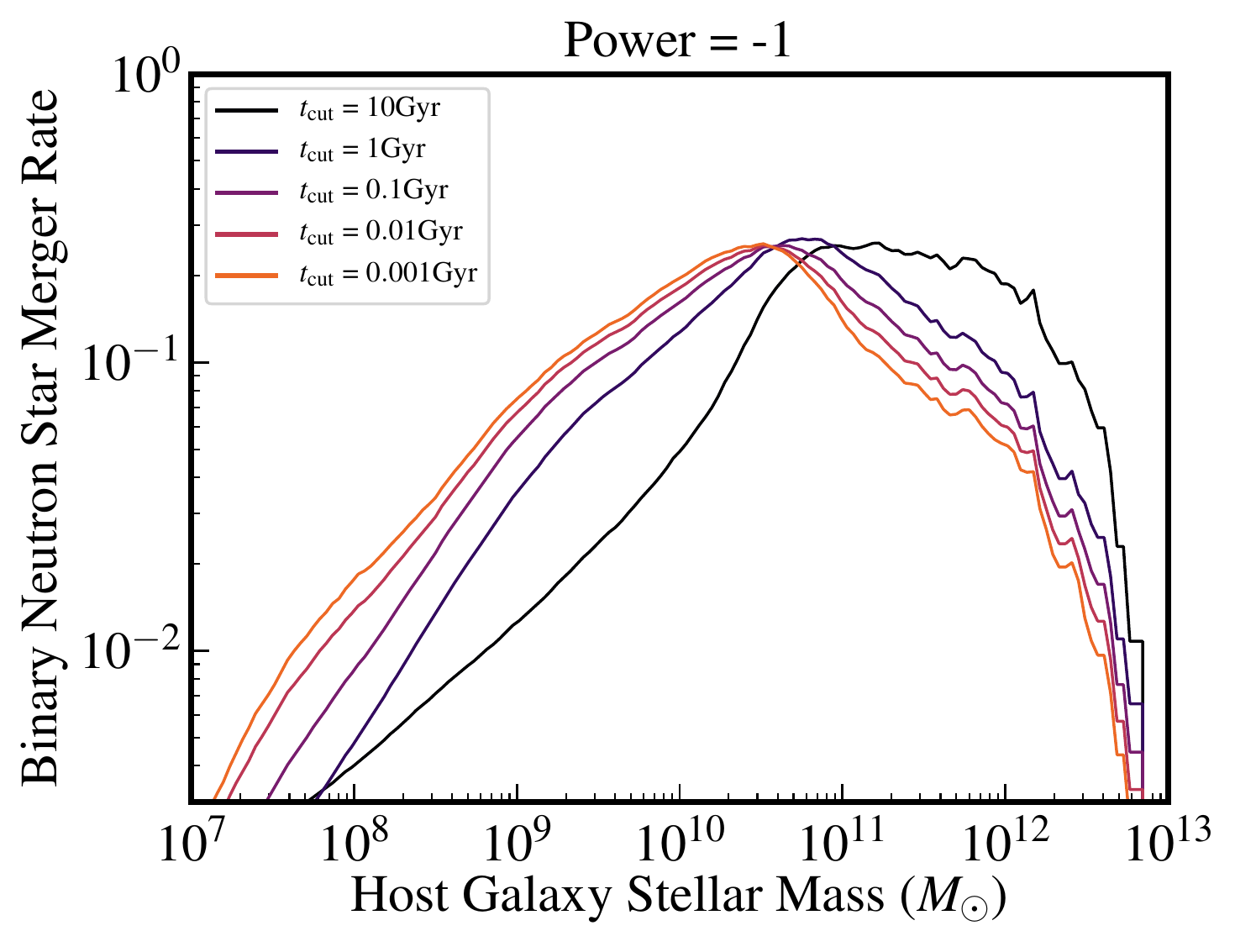}
    \caption{The individually normalized BNS merger rates as a function of stellar mass for varied power law exponents (left) and varied $t_{\mathrm{cut}}$ values (right).  There is significant variation in the predicted host galaxy mass functions when the DTD is perturbed from the fiducial values.}
    \label{fig:spread}
\end{figure*}

Figure \ref{fig:spread} shows the host galaxy mass functions for the power law DTDs with varied power law exponents and $t_{\mathrm{cut}}$. 
For completeness, we explore a large range of values for the power law exponents and cutoff times $t_{\mathrm{cut}}$ which go beyond what is believed to be physically correct~\citep{2017Cote}. 
These values are included to demonstrate the variation in host galaxy mass functions which would result from varied DTDs. 
Even with this large spread of DTDs, we find that the results shown in Figure \ref{fig:fiducial} are broadly stable. 
In particular, despite the very significant variation in the DTDs, all cases show a peak BNS merger rate that occurs in galaxies with a stellar mass in the range $10^{10}$-$10^{11}$ $M_\odot$.

The DTD for which $s=0$ is of particular interest in Figure \ref{fig:spread} because it tracks the total stellar mass found in each mass bin -- nearly independent of star formation history.\footnote{There is a dependence on the amount of stellar mass that formed in the past $t_{\mathrm{cut}} = 10 \mathrm{Myrs}$, but this is expected to be only a $\sim0.1\%$ correction.} 
Owing to the shape of the simulated (and observed) galaxy stellar mass functions, the peak of the stellar mass distribution is in galaxies with stellar masses between $10^{10}$ and $10^{11}$ $M_\odot$.
Therefore, the majority of BNS mergers for this DTD are also found in that mass range. 
Importantly, because the predicted host galaxy mass function for a DTD with power law index of $s=0$ is nearly independent of galactic formation history, this result is not very sensitive to the detailed SFRHs predicted by IllustrisTNG, but only the shape of the galaxy stellar mass function.
Insofar as other simulations or models reproduce the same galaxy stellar mass functions, their predicted host galaxy mass function for $s=0$ would be nearly identical.

Changing the power law exponent to values away from $s=0$ introduces a direct dependence on the assumed star formation history by placing emphasis either on the older or younger stellar populations.
Specifically, power law exponents higher than $s=0$ lead to systematic changes in which the host galaxy mass function is biased toward galaxies with older stellar masses.
This naturally results in a shift of the peak in the host galaxy mass function toward older, more massive systems.
Conversely, changing the power law exponent to values lower than $s=0$ (which is the more physical case) biases the host galaxy mass function toward systems with younger stellar populations.  
Thus, as the power law exponent is decreased, there is an expectation that an increasing number of BNS mergers occur in low mass galaxies with current or recent ongoing star formation.
For a fixed DTD, the detailed shape that we predict for the BNS merger host galaxy mass function is dependent on the IllustrisTNG galaxy stellar mass function and SFRHs, and therefore should be checked against other models (e.g. EAGLE, SIMBA, etc.).
However, owing to observational constraints provided by the cosmic star formation rate density and redshift dependent galaxy stellar mass functions, we do not expect these results to substantially change.

Despite the stability in the the peak of the host mass function across different DTDs, there is still significant spread in the resulting host mass functions at other masses.
For example, when examining the BNS merger rate in galaxies with a host mass near $10^9$, the BNS merger rates differ by 1-dex between the $s=2$ and $s=-2$ DTDs with a fixed $t_{\mathrm{cut}}$. 
There is an even greater spread in the highest mass systems where the merger rate differs by 2-dex between the $s=2$ and $s=-2$ DTDs. 
A similar range in host mass functions occurs across the different $t_{\mathrm{cut}}$ values at a fixed value of $s$. 
The predicted variability in host galaxy mass functions suggests that as GW BNS detections with EM follow-up observations mount, a careful comparisons of observed and predicted host galaxy mass functions could be used to constrain the true DTD for BNS mergers.
These results agree with \cite{Safarzadeh2019} and running KS-tests on our power law host mass functions also results in $\mathcal{O}(1000)$ observations being required to determine a true DTD.

\begin{figure*}
	\includegraphics[width=\columnwidth]{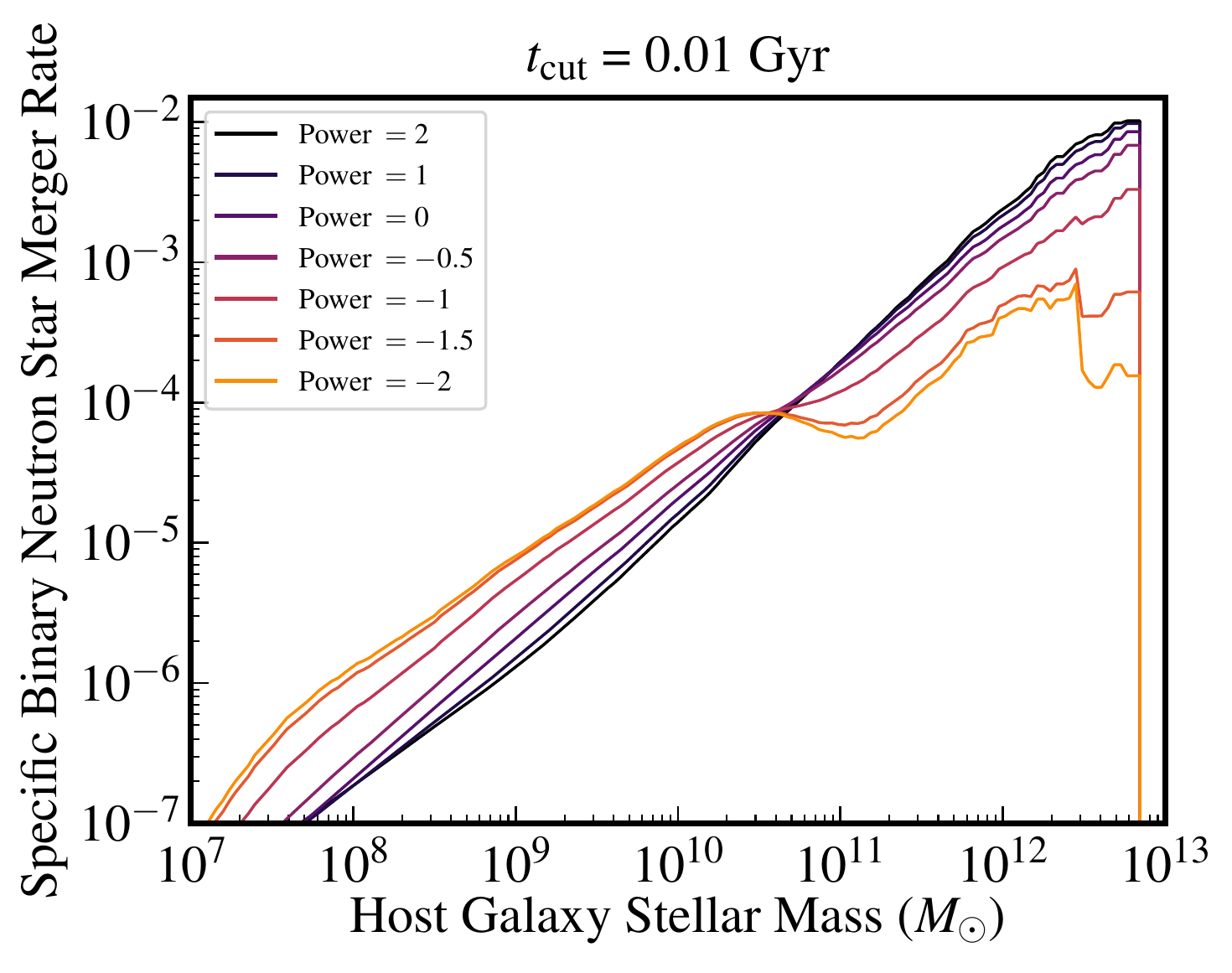}
	\includegraphics[width=\columnwidth]{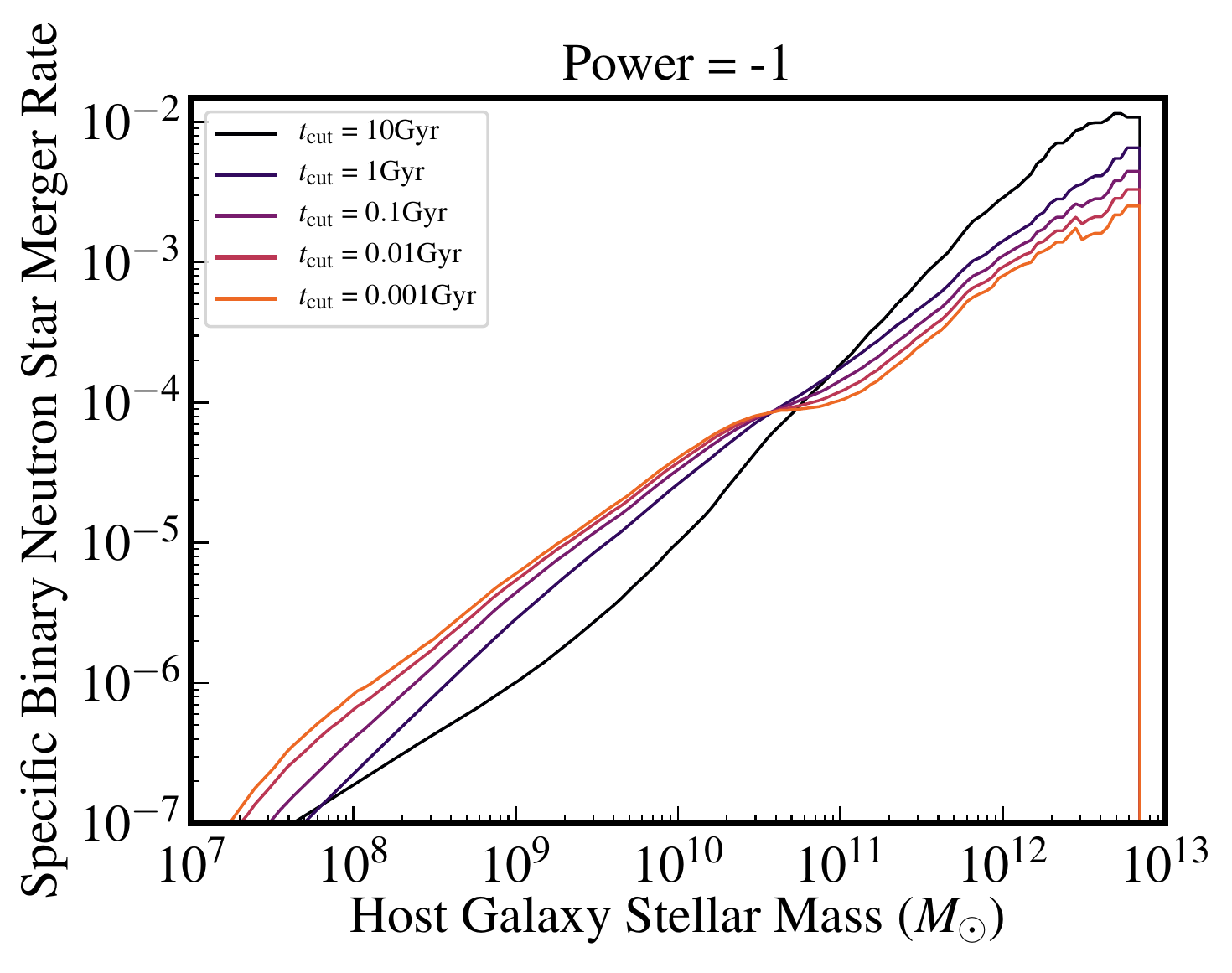}
    \caption{
The individually normalized BNS merger rates as a function of stellar mass per galaxy for varied power law exponents (left) and varied $t_{\mathrm{cut}}$ values (right).  While the host galaxy mass function (Figure~\ref{fig:spread}) predicts most BNS mergers will happen in roughly Milky Way mass galaxies when averaged over the whole galaxy population, this host galaxy specific-mass function (this figure) indicates that the rate of BNS merger rate is higher is higher in higher mass galaxies, when compared on an individual basis.}
    \label{fig:specific}
\end{figure*}

When optimizing BNS merger event follow-up, a question arises of which galaxies should be targeted first.  
While Figures~\ref{fig:fiducial} and~\ref{fig:spread} indicate that most BNS mergers will occur in roughly Milky Way mass galaxies,  
Figure \ref{fig:specific} shows the host mass function normalized by the number of galaxies in each mass bin which indicates the predicted number of BNS mergers per galaxy.
Here, the host mass function no longer peaks in the $10^{10}$-$10^{11}$ $M_\odot$ range, but instead peaks at larger masses (in the $10^{12}$-$10^{13}$ $M_\odot$ range). 
This indicates that while our analysis predicts that most BNS merges will occur in $\sim$ Milky Way mass galaxies, the highest likelihood of finding a BNS merger based on a single observation still favors more massive systems, simply because they have more mass.
This conclusion is somewhat dependent on the detailed assigned prescription for the BNS merger DTD.
In particular, while the fiducial values (see the magenta line in the left panel of Figure~\ref{fig:specific}) clearly peaks at the highest mass bin resolved in the IllustrisTNG volume, the steeper exponent cases ($s=-1.5$ and $s=-2$) are much flatter above $M_*=10^{10.5} M_\odot$.

A closer examination of how the BNS merger rate correlates with different observables indicates a dependence on the DTD. 
This analysis was conducted by comparing, through the Pearson correlation coefficient, the BNS merger rate for each galaxy to one of three observable properties: its star formation rate (SFR), its blue luminosity, and its stellar mass. 
For very steep and negative DTDs, $s=-2$, we find that SFR is best correlated with merger rate (R=0.978). 
For less steep and negative DTDs, $s=-1$, we find that blue luminosity is best correlated (R=0.993). 
For flat or increasing DTDs, $s=0+$ we find that stellar mass is best correlated (R=1.0, 0.994, 0.985 for s=0,1,2 respectively).

\section{Discussion}
\label{sec:Discussion}

The ability to connect LIGO-detected BNS merger events to their host galaxy opens new scientific opportunities.
Specifically, while transient event detection and host galaxy association is well-established in astronomy, traditional methods for kilonova detection yield little direct information about the progenitor system.
In contrast, wave form fitting of LIGO detected compact object merger events provides detailed information about the progenitor system including the masses of the merging objects.
This new information links mergers of specific object types to host galaxies with a limited level of ambiguity or uncertainty that was not previously possible.
As we have discussed in this paper, this opens up the possibility of developing a more intimate link between galactic star formation rate histories (SFRHs), BNS delay time distributions, and the observed host galaxy stellar mass function.

In this paper, we have leveraged the galactic SFRHs from the IllustrisTNG cosmological galaxy formation model. 
At some level, these SFRHs are likely not a perfect reflection of real galactic SFRHs. 
However, the model is able to broadly match the cosmic star formation rate history as well as the redshift dependent galaxy stellar mass function.
This gives us a reasonable level of confidence that these simulated SFRHs provide good approximations to those of real galaxies.  
Moreover, since fairly different physical models (e.g. those of Illustris model, the EAGLE model, and semi-analytical models) yield similar BNS merger rates, 
we believe the more holistic analysis obtained from the TNG-100 simulation can further our understanding of the host galaxy mass function and its dependence on the DTD.

An important feature of the IllustrisTNG simulation is that it self-consistently tracks gas- and stellar-phase metallicities.
The stellar metallicities, in turn, impact the results of the metallicity-dependent DTDs, such as the BPASS DTD.
It has been shown that IllustrisTNG's stellar metallicity vs stellar mass relation generally agrees with observations but is too flat leading to higher metallicities at lower masses \citep{2018Nelson}.
The largest discrepancy is $\sim 0.5$ dex near $10^{10.5} \; M_\odot$.
To understand how this uncertainty affects our results, we use two host galaxy mass functions with the BPASS DTD. 
Each host galaxy mass function is made by setting all stellar metallicities to either $Z=0.0001$ or $Z=0.014$.
Given the large difference in metallicity, $\sim 2.5$ dex, between these host galaxy mass functions, we expect any differences to be larger than those introduced from uncertainties in the IllustrisTNG stellar metallicities.
We find very little variation between the host galaxy mass functions between the lowest and highest metallicities across all galaxy masses.
Given the large spread in metallicities used in creating these DTDs, it is unlikely that our results would be changed significantly if we instead employed stellar metallicities from a different galaxy formation model.
Given that our results do not change when accounting for the more complicated BPASS DTD, additional credibility can be given to results derived from power law DTDs.

Some efforts have already begun to explore the new connection between merger events and their host galaxy using population synthesis codes and various star formation histories.
These studies investigate how the host galaxy mass function is affected by the merger progenitors.
The unique contribution of this paper is to focus on systematic variations of the employed DTD coupled to cosmologically motivated SFRHs.
A similar set of DTD variations was employed in \cite{Safarzadeh2019}, albeit with analytically simplified SFRHs.
They concluded that the host galaxy mass function peaks at high masses and that $\mathcal{O}(10^3)$ observations are required to constrain the a true power law distribution.
This paper agrees with their conclusions for the host galaxy mass range that they cover, $10^9 - 10^{11.25} M_\odot$. 
This study extends \cite{Safarzadeh2019} by pairing a broad set of assumed DTDs to SFRHs naturally derived in a cosmological environment and examining how the assumed DTD affects the host galaxy mass function over the large range of host masses allowed by IllustrisTNG.

Similar results to those presented in this paper have also been discussed in \citet{2019ArtaleMass, 2019Safarzadeh, 2020McCarthy}.
\cite{2019ArtaleMass} uses the EAGLE simulation to create a stellar mass vs specific BNS merger rate plot similar to Figure \ref{fig:specific}.
They find that stellar mass is an excellent tracer for specific merger rate.
This result is consistent with our result up to $\sim 10^{10.5} M_\odot$.
However, at higher masses we find a dependence on the DTD, causing faster merging times to not depend on stellar mass.

\cite{2019Safarzadeh, 2020Adhikari, 2020McCarthy} also present host galaxy mass functions using different SFRH models.
\cite{Safarzadeh2019} uses an analytic model with the set us DTDs used in this paper to understand how the host galaxy mass function is affected by the DTD.
Our results are generally consistent with theirs, but our extended range of host masses allow us to see that most BNS mergers do not happen in galaxies with highest mass, but in galaxies with masses between $10^{10}$ and $10^{11} M_\odot$.
\cite{2020McCarthy} also uses an analytic model but paired with SDSS observations to explore the host galaxy mass function along with other host observables.
For the host galaxy mass function, our results are consistent with theirs.
However, our larger range of DTDs presented in Figure \ref{fig:spread} reveal the large spread between different assumed DTDs and the stable peak near $10^{10.5} M_\odot$.
\cite{2020Adhikari} also find that other host observables paired with stellar mass are necessary to obtain a better understanding where BNS merge.
Overall, we find that future explorations of this topic will need to consider a wide range of DTDs and the full range of the observable in question.

While the work we present here continues our understanding of what we can learn from observations of BNS host galaxies, further investigations are necessary to fully understand how BNS form and evolve.
One example of such an investigation is to expand the set of DTDs examined using the methods in this paper.
The set of DTDs we examine are broad, covering those most commonly referenced~\citep[e.g.][]{Safarzadeh2019, Eldridge2016}, but we do not exhaustively search the full range of DTDs proposed ~\citep[e.g.][]{2019Simonetti, 2012Dominik}.
Also, our convolution of IllustrisTNG's SFRH with our DTDs does not include any form of natal kicks. 
If these kicks are strong enough to dislodge the binary from smaller galaxies, it is possible their addition would weight the host mass functions toward higher mass galaxies.
With a greater range of DTDs examined and a more detailed convolution, we will gain a clearer picture of where BNS mergers are located, which delay times can be distinguished using the host galaxy mass function, and the most likely places they will be observed. 
Another way to incorporate a more complete set of DTDs would be to use a varied set of population synthesis models which cover a wide range of binary separations, kick velocities, initial mass functions, etc.

Including other star formation histories could also provide a more detailed look at the spread in possible host galaxy mass functions. 
While IllustrisTNG is broadly consistent with the cosmic star formation rate density and redshfit dependent galaxy stellar mass functions \citep{Pillepich2018b}, its accuracy should not be over interpreted and different simulations will surely produce somewhat varied star formation histories that could impact our results. 
However, we can say that up to their mass cutoff, our results align with \cite{2019ArtaleMass} who found no significant difference when comparing results from Illustris and Eagle. 
The lack of variation in \citet{2019ArtaleMass} most likely indicates that -- while there is some variation -- the SFRHs in Illustris and EAGLE are sufficiently similar to not significantly impact the results.
Thus, by adopting the SFRHs from galaxy formation simulations and assumptions about the functional form of the DTD, predictions can be made about the BNS host galaxy mass function.  
Additionally, similarities between the different simulations suggest that uncertainties in the poorly constrained DTDs are likely larger than the uncertainties introduced from the SFRHs. 
In particular, the detailed shape of the BNS host galaxy mass function will be sensitive to assumptions about the DTD.

\section{Conclusion}
\label{sec:Conclusions}

We presented predictions for the host galaxy mass function and host galaxy specific-mass function for BNS mergers.
Our predictions were generated by convolving a set of power law and BPASS DTDs with the star formation histories from the IllustrisTNG cosmological simulation. Our main conclusions are as follows:

\begin{enumerate}
    \item We find almost no difference between the host galaxy mass functions produced by our fiducial power law (slope of $s=-1$, minimum time of $t_{\rm cut}=10^7$ yrs) and the BPASS DTDs (Figure~\ref{fig:fiducial}).
    \item The peak of the host galaxy mass function occurs around the Milky Way mass scale, with roughly $\sim50\%$ of BNS mergers happening in the $10^{10} M_\odot < M_* <  10^{11} M_\odot$ mass range for our fiducial DTDs (Figure~\ref{fig:fiducial}). This mass bin includes NGC 4993, the host galaxy of GW170817.
    \item While the detailed shape of the host galaxy mass function is sensitive to details of the adopted DTD, the peak does not change significantly when varying over a broad range of DTDs (Figure~\ref{fig:spread}). The peak of the host galaxy specific-mass function is similarly insensitive to changes in the adopted DTD (Figure~\ref{fig:specific}).
    \item The peak of the host galaxy specific-mass function is located in the highest mass bin for the fiducial power law DTD and BPASS model. Thus, while we expect most BNS mergers to happen in somewhat lower mass systems for our fiducial DTDs, high mass galaxies are more likely to host a BNS merger on a per-galaxy basis (Figures~\ref{fig:fiducial} and~\ref{fig:specific}).  
    \item Host galaxy mass functions constructed from different DTDs vary up to one dex at low masses and up to two dex at high masses.  This provides an opportunity through which an observationally reconstructed host galaxy mass function can be used to constrain the true BNS DTD (Figure~\ref{fig:spread}).
    \item The observable galactic property (or properties) that is expected to provide the best correlation with the BNS merger rate depends on the true DTD.
\end{enumerate}

In the short term, the results found here in both the host mass and specific-mass functions paint an interesting picture on how astronomers should structure electromagnetic follow-ups for BNS events. 
The peak of the host galaxy specific-mass function laying in the highest mass bin suggests that the optimal way to quickly find the resulting kilonova from a BNS merger would be to search the highest mass galaxies first. 
This agrees with the current method most follow-up strategies use in locating BNS mergers ~\citep[e.g.][]{2016Gehrels, 2017Arcavi, 2016Singer}. 
However, the peak of the host mass function laying in the mass range $M_*=10^{10} - 10^{11} \; M_\odot$  suggests that this method will miss, or take a longer to locate, most of the BNS mergers. 
Determining the true DTD would allow for more efficient electromagnetic follow-up by determining which observable: SFR, blue luminosity, or stellar mass, best correlates with BNS merger rate.

In the long term, LIGO/Virgo/KAGRA will create a host mass function which can be used to determine the true BNS DTD. 
With this DTD, the minimum delay time, $t_{\rm cut}$, can constrain the proportion of BNS systems which form through highly eccentric and low separation fast-merging channels. 
Understanding this proportion will place constraints on natal kick velocity and common envelop efficiency. 
The minimum delay time can also determine whether BNS mergers are the dominant source of r-process elements. 
The overall shape of the true DTD allows various physical parameters of BNS systems to be constrained, such as the progenitor's metallicity, masses, mass ratio, common envelope efficiency, natal kicks, and initial binary separation through comparisons with resulting DTDs from population synthesis codes.

\section*{Acknowledgements}
The authors thank Steve Eikenberry for his useful ideas and comments.
JCR acknowledges support from the University of Florida Graduate School's Graduate Research Fellowship.
PT acknowledges support from NSF grant AST-1909933, NASA ATP Grant 19-ATP19-0031. IB acknowledges support from the Alfred P. Sloan Foundation and the University of Florida.

%\bibliography{citations}{}
%\bibliographystyle{aasjournal}

\end{document}